\documentclass[letterpaper]{JHEP} 

\usepackage{graphicx}
\usepackage{psfrag}
\usepackage{url}

\providecommand{\MSbar }{\ensuremath{ \overline{\rm MS} }}
\newcommand{\xbj}{\ensuremath{x_{\rm bj}}}
\newcommand{\ktcut}{\ensuremath{k_{t~\rm cut}}}

\makeatletter  
\preprint{hep-ph/0212053\\
\@date
}
\makeatother

\title{ Rapidity and Transverse Momentum Distributions of DIS
        1-jet Inclusive Cross Section \\ --- A Next-to-Leading
        Order Monte-Carlo Prediction}

\author{John Collins$^1$\hspace{10pt} 
        Xiaomin Zu$^2$ 
        \\
        Physics Department,
        Penn State University, \\
        104 Davey Laboratory,
        University Park PA 16802,
        U.S.A. 
\\ \email{$^1$ collins@phys.psu.edu}
\\ \email{$^2$ xiaozu@phys.psu.edu}
}

\date{3 December 2002}


\abstract{We have implemented in the RAPGAP program a previously
  derived subtraction method for next-to-leading order (NLO)
  corrections in Monte-Carlo (MC) event generators, and we show
  results for jet production in deep inelastic scattering.  At small
  $\xbj$, NLO corrections are comparable to the LO cross section,
  because of the large gluon density.  We devise a jet observable that
  is particularly sensitive to the treatment of parton kinematics,
  which our method is intended to treat correctly.  We compare the
  results of the calculation with LO calculations and with the
  previously existing treatment of NLO processes in RAPGAP.
  Substantial corrections to the event distribution are found.  }

\begin{document}

%==========================================================
\section{Introduction}
\label{sec:intro}

In contrast to ``fixed order'' calculations, Monte-Carlo (MC) 
event generators attempt to predict the bulk of the 
exclusive components of perturbative QCD subprocesses and to give the
details of the final states. While this is at the expense of 
model dependence in the non-perturbative hadronization, it is also 
the case that MC event generators incorporate a useful approximation
to the resummation of large logarithms in their parton showering.
This provides a successful phenomenology of infrared sensitive 
cross sections.

Now, it is quite hard to treat non-leading corrections in MC
event generators.  However, this is also quite essential. 
For example, in deep inelastic scattering (DIS) at small $\xbj$
or in diffractive DIS, the gluon density is substantially larger 
than the quark densities, and the next-to-leading order (NLO)
gluon-induced process is comparable to the leading order (LO)
quark-induced process. Therefore, to get a sensible phenomenology,
the gluon-induced NLO corrections must be included.

Precisely because MC event generators treat exclusive components of
the cross section, standard ``fixed order'' calculations of hard
scattering coefficients are not well adapted for use in event
generators.  Some symptoms of this are that in LO, the coefficients
contain $\delta$-functions of parton kinematics, and that in NLO the
coefficients are singular distributions (generalized functions).  They
only give results that can be compared with hadronic level cross
sections after an integration over a wide range of parton kinematics,
which provides a kind of infrared safety. Standard fixed-order
calculations approximate the partons as on-shell with zero transverse
momentum.  For cross sections that do not include a wide integration
over parton kinematics, only a correct treatment of parton kinematics
allows a correct conversion of the generalized functions to ordinary
functions of kinematic variables.  In fact, the standard
approximations used in hard-scattering coefficients cause
complications and difficulties in the derivation of the algorithms
used in event generators.

In this paper, we will examine the rapidity and transverse momentum 
distribution of the 1-jet inclusive cross section in DIS.
Even though this is a jet cross section, it is not infra-red safe
because its LO hard scattering coefficient is a $\delta$-function, and in the
approximation that parton virtuality and transverse momentum are
ignored, the jet momentum can be calculated from the measured
4-momentum of the scattered electron.  Thus the $\delta$-function
appears in the hadron-level jet cross section, in this simple
parton-model approximation. 
Similarly, the conventional NLO corrections
are singular at the parton-model values. Because we only integrate over
a narrow range of parton kinematics around these singularities, 
it is not clear how to apply these conventional NLO hard
scattering coefficients to this cross section. However, it is crucial for a 
good NLO MC method to consistently treat this kind of cross section with 
NLO accuracy.

As we argued in \cite{phi3,JC,zu}, a 
proper treatment of higher order (NLO and beyond) corrections in 
event generators requires (a) that the parton kinematics should be 
treated more exactly and (b) that the hard scattering coefficients should be
ordinary functions, not generalized functions.  
This is in contrast to other
proposals \cite{other} for matrix element corrections to event generators.
In \cite{phi3} our approach is proposed
in general and
is designed to be generalizable to all orders in perturbation theory.

More concretely, in \cite{JC} this method was applied to DIS in a form
specifically designed for use in 
event generators using the Bengtsson--Sj{\"o}strand \cite{BS} algorithm
for showering.  Such event generators include
RAPGAP \cite{jung} and PYTHIA \cite{pythia}.
We have now implemented\footnote{Only gluon-induced NLO corrections are 
implemented, as in \cite{JC}.  The quark-induced NLO terms have the
complication of a soft 
divergence, which we plan to deal with in a later paper.  
Gluon-induced NLO corrections are important in DIS, because they are
not suppressed compared to the LO terms whenever the gluon
distribution is large.
}
this method 
in RAPGAP, a widely used event generator 
for both inclusive and diffractive DIS. The implementation is based on
earlier work by Schilling \cite{sabine}.

In this paper we show the results of the 1-jet inclusive cross 
section calculated by our subtraction method. We also compare
our method with a method previously implemented in  RAPGAP, which uses 
a $k_t$-cut applied to the simple unsubtracted NLO cross
section. The $k_t$-cut method is intended mainly for the  
di-jet rate calculation of high $p_t$ jet events. 

We also compare our results to those of a LO calculation from 
RAPGAP, where the $\delta$-function in the 1-jet inclusive cross
section is smeared out only by the parton showering and the hadronization. 
At small $\xbj$ the gluon-induced NLO contributions are enhanced, and 
we expect the cross section with NLO accuracy to be a very broad distribution,
and the three methods should give very different results. At large $\xbj$
the NLO corrections are suppressed by $O(\alpha_s)$, we expect to see 
similar results by these methods.

The organization of this paper is the following. In Sec.\ \ref{sec:nlo} we
briefly describe  the subtraction method and the $k_t$-cut method in RAPGAP.
In Sec.\ 
\ref{sec:Xsec} we show and discuss the results of the 1-jet inclusive cross 
sections for DIS at small $\xbj$ and large $\xbj$ predicted by the 
three different methods. In Sec.\ \ref{sec:concl}, we discuss the possible 
applications and future work of the NLO RAPGAP with subtraction method.

%==========================================================
\section{NLO hard scattering}
\label{sec:nlo}

In \cite{JC}, we devised a subtraction method for NLO corrections, and 
obtained the coefficient function for the theoretically simple
yet phenomenologically significant gluon-induced correction to DIS.
It was assumed that the
Bengtsson-Sj{\"o}strand (BS) \cite{BS} algorithm is used
for the initial state parton shower, as in RAPGAP. A point-by-point
subtraction  
in the phase space  ensures that the hard-scattering coefficient is
non-singular \cite{JC}.  The pdfs needed are specific to the showering 
algorithm \cite{MW}, and those needed to match the BS algorithm were
calculated in \cite{zu}.
The quark-induced NLO corrections have the complication of a soft
divergence, and we plan to treat this case, matched to the BS
algorithm, in a later paper; the quark-induced term is a genuine
correction, order $\alpha_s$ compared with the LO process, whereas the
gluon-induced term is not suppressed at all when the gluon density is large.

In RAPGAP, the previously implemented method for NLO was the $k_t$-cut
method, designed mainly to  
give the correct di-jet rate at large $Q^2$. Here the unsubtracted parton-level 
matrix elements of the NLO processes are used with a cut on the 
transverse momentum $k_t$ of the out going partons in the center of 
mass frame of $\gamma^* + {\rm parton}$ scattering. $\ktcut$ is required to be 
large enough so that the NLO contributions are smaller than the 
total cross section. Usually $\ktcut^2 = 4\,{\rm GeV}^2$. This gives us a 
large logarithm $\log (\ktcut^2/Q^2) $ in the integrated NLO cross
section when $Q^2$ is large.

%----------------------------------------------------------
\section{Rapidity and Transverse Momentum Distribution of DIS 1-jet 
Inclusive Cross Section}
\label{sec:Xsec}

\subsection{Theoretical and experimental considerations}
To probe the differences between different treatments of NLO
corrections in an event generator, we need an infrared 
sensitive cross section that potentially gives us the maximum
difference between the different methods. We choose the rapidity and
transverse 
momentum distribution of the 1-jet inclusive cross section, 
$d^2\sigma/(d\hat{\eta}dx_t)$ in the laboratory frame, with the rapidity and 
transverse momentum measured relative to the simplest parton-model
values.  At the LO parton level, the 
scattered quark has a transverse momentum of $E_t^{\rm PM}$ and
pseudorapidity $\eta^{\rm PM}$, both of which can be calculated from the 
momentum of the scattered electron, and so we define
\begin{eqnarray}
  x_t &\equiv& \frac{E_t^{\rm jet}}{E_t^{\rm PM}},
\\  
 \hat{\eta}&\equiv&\eta^{\rm jet} - \eta^{\rm PM}.
\end{eqnarray}
If we ignore parton virtuality and transverse momentum, the LO
cross section is a simple $\delta$-function 
\begin{equation}
  \frac{d^2\sigma}{d\hat{\eta}dx_t}  \propto  \delta(x_t-1) \delta(\hat{\eta}).
\end{equation}
In real QCD, with effects from higher order QCD subprocesses and the 
hadronization, the $\delta$-function will be smeared out. The shape of 
the smeared $\delta$-function predicted by a MC event generator will provide 
us information about its treatment of the NLO corrections.

We calculate the cross section from the event generator RAPGAP by
three methods: 
\begin{enumerate}
\item LO hard scattering, where jet structure is generated exclusively 
  by parton showering and hadronization.
\item With NLO corrections implemented by the $\ktcut$ method.
\item With NLO corrections implemented by our subtraction method.
\end{enumerate}
Due to the enhanced gluon density at small $\xbj$, the cross section
in the small $\xbj$ region is likely to give us the maximum 
difference between the different methods. At large $\xbj$ the 
NLO corrections are suppressed by $\alpha_s$, and therefore in
this region the cross section is not sensitive to the difference
between the NLO 
methods. So in this paper we calculate this cross section in 
two different regions, $10^{-4} < \xbj < 10^{-2}$, where we expect to see
very different results from the three calculations, and $0.02 < \xbj < 1$,
where similar results are expected. 

Our calculations are performed under the following conditions, with
hadron kinematics appropriate to the collider experiments at HERA:
\begin{itemize}
\item The positron is moving in the $-z$ direction with $P_z^e = -27.5$ GeV.
\item The proton is moving in  the $+z$ direction with $P_z^p = 820$ GeV.
\item $20 < Q^2 < 50$ GeV$^2$.
\item In LO and $k_t$-cut calculations, the LO pdfs Cteq4L are used.
\item In RAPGAP with our subtraction method, we use the BS-algorithm
  specific pdf calculated by a scheme change \cite{zu} from the
  $\MSbar$ pdf, Cteq4M. 
\item The jet-finder is PXCONE, with $E_t$-mode and cone radius$=1$.
\item The running scale of the pdfs and $\alpha_s$ is $Q$.
\item Initial state showering is turned on, with $Q^2$ ordering.
\item Final state showering is on.
\item Hadronization and proton remnant are required.
\item The LO subprocess and (for the NLO calculations)
      the gluon-induced NLO subprocesses are chosen.
\item Only light quarks are considered.
\end{itemize}
Figs.\ \ref{fig:graph2d}, \ref{fig:smallxeta} and \ref{fig:smallx} give the
results for $10^{-4} < \xbj < 10^{-2}$, while Figs.\
\ref{fig:largexeta} and 
\ref{fig:largex} give the results for $0.02 < \xbj < 1$.

%==========================================================
\subsection{Cross Section at small $\xbj$}
\label{sec:sx}
\FIGURE{
\centering
\includegraphics*[1in, 5.3in][7in, 7.6in]{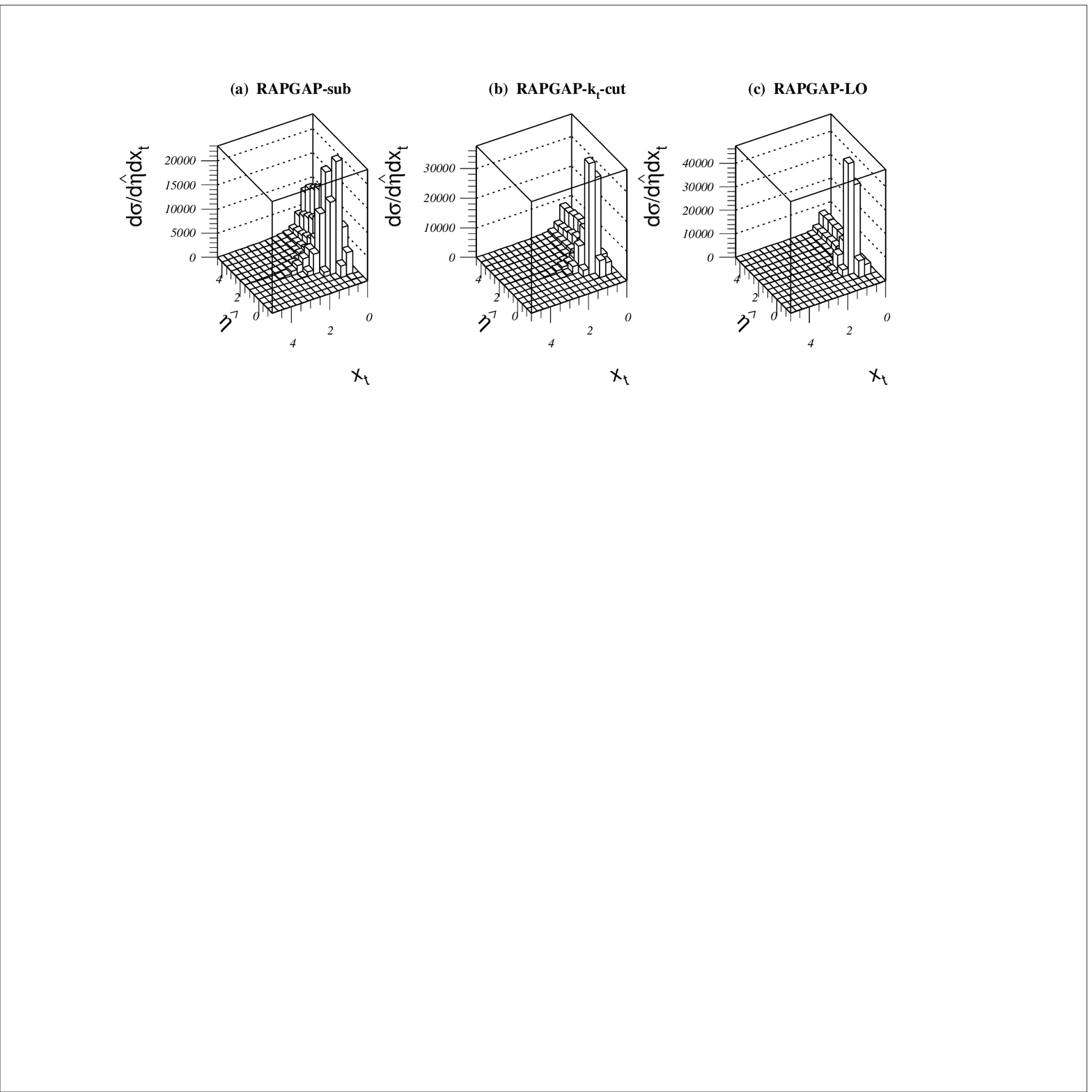}
\caption{Transverse momentum and rapidity distribution of 1-jet inclusive 
cross section for $10^{-4} < \xbj < 10^{-2}$. (a) RAPGAP with the subtraction
method, (b) RAPGAP with the $k_t$-cut method and (c) LO RAPGAP. Note that the
vertical scales for these three graphs are different.}

\label{fig:graph2d}
}

In Fig.\ \ref{fig:graph2d} we plot $d^2\sigma/(d\hat{\eta}dx_t)$ as a 
function of $x_t$ and $\hat{\eta}$ for the small $\xbj$ region, 
$10^{-4} < \xbj < 10^{-2}$.  It clearly shows that the  
LO (graph c) prediction is strongly peaked at the PM value 
$x_t = 1$ and $\hat{\eta} = 0$, as expected.
The cross section 
predicted by the $k_t$-cut method (graph b) is somewhat broader but still 
peaked at the PM values.
In contrast, the prediction from our subtraction method 
(graph a) is broadly distributed; although it has a maximum
close to the PM position, the peak is not at all $\delta$-function like.
This indicates that when the gluonic NLO term is comparable to the LO
term, a correct treatment of off-shell parton kinematics has a
substantial effect on the shape of the cross section.

\FIGURE{
\centering
\includegraphics*[1in, 3.2in][7in,7.6in]{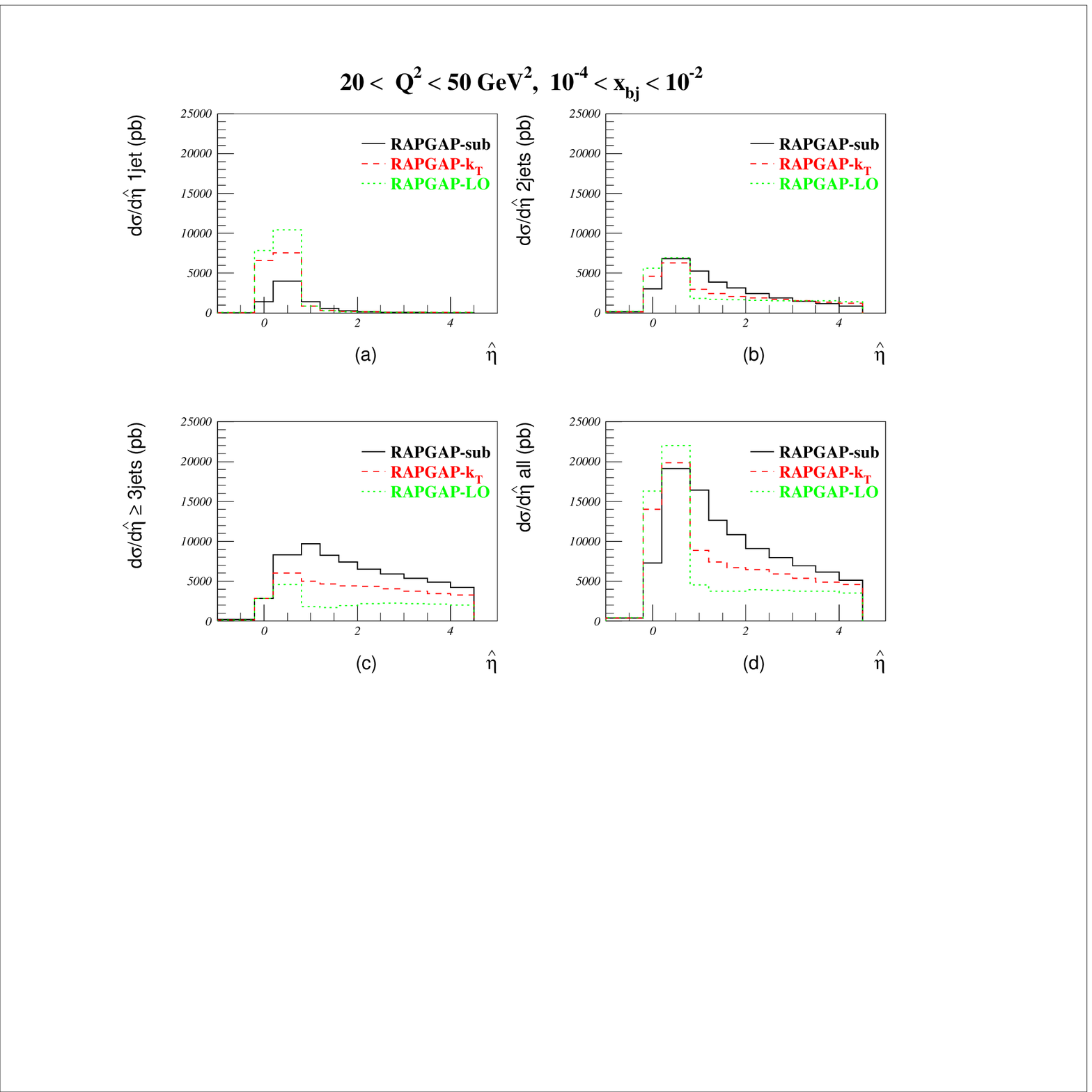}
\caption{Rapidity distribution of 1-jet inclusive cross section for
 $10^{-4} < \xbj < 10^{-2}$.  The four graphs show the following contributions:
(a) 1-jet events, (b) 2-jet events, (c) $\geq3$-jet events, (d) 
total cross section.  The solid curve, dashed curve and dotted curve are
predictions from the subtraction method, the $k_t$-cut method and the LO 
RAPGAP, respectively.}

\label{fig:smallxeta}
}

For a more quantitative comparison between these methods, we 
now concentrate on the one parameter cross sections
$d\sigma/d\hat{\eta}$ and $d\sigma/dx_t$, shown in 
Figs.\ \ref{fig:smallxeta} and \ref{fig:smallx}. 
In each figure, we show in graphs (a), (b) and (c) the components of the
1-jet inclusive cross section separated according to the number of
jets in the final state: 1 jet, 2 jets and 
$\geq3$ jets. Then in graph (d), we give the total 1-jet inclusive cross 
section.  The 
solid curves, dashed curves and the dotted curves  are 
the predictions of our subtraction method, the $k_t$-cut method, and
the LO calculation, 
respectively. 

\FIGURE{
\centering
\includegraphics*[1in, 3.2in][7in,7.6in]{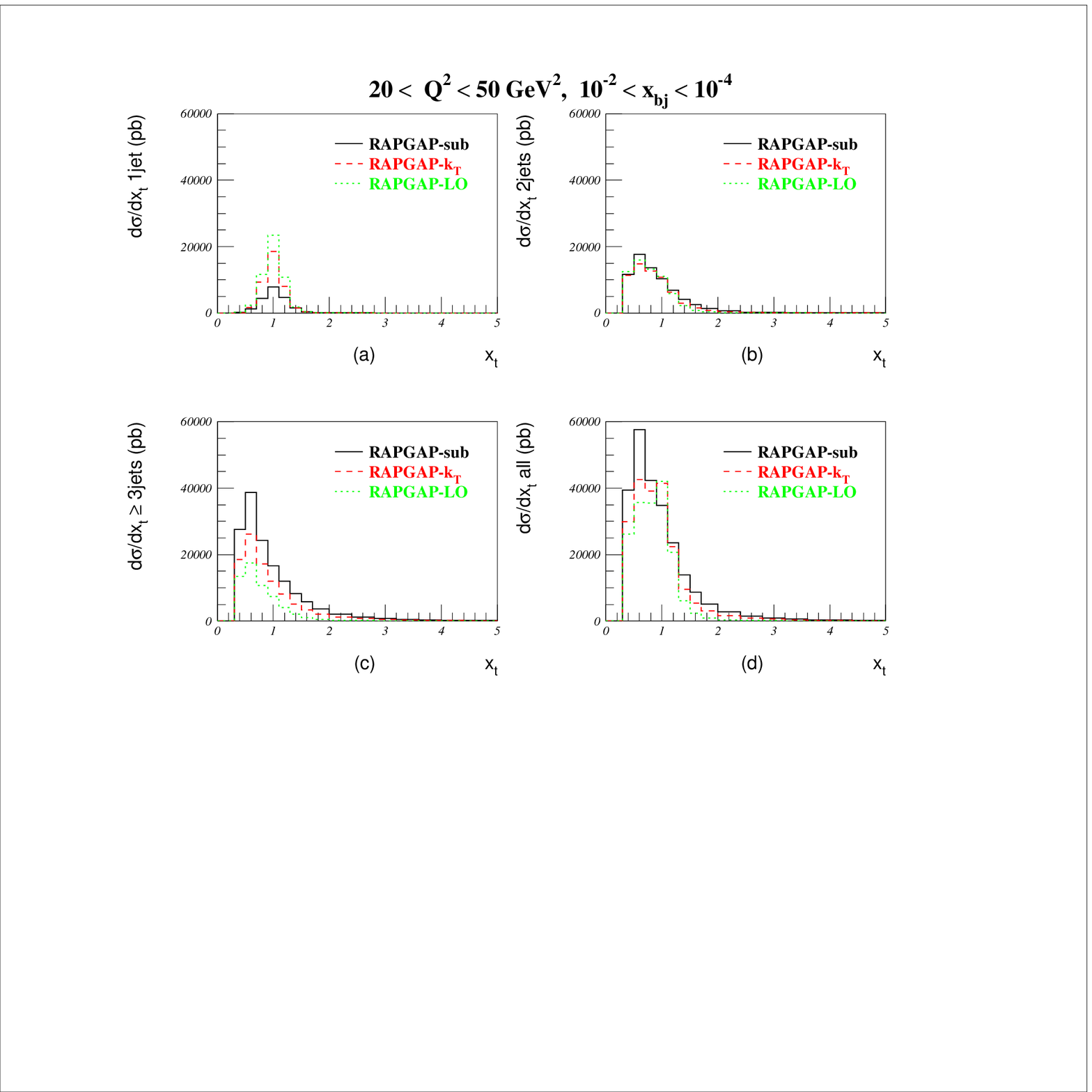}
\caption{Transverse momentum distribution of 1-jet inclusive cross
  section.  The coding of the curves is the same as in Fig.\
  \ref{fig:smallxeta}.} 
\label{fig:smallx}
}

The differences between the predictions are evident. The subtraction
method gives the broadest distribution, while the $k_t$-cut method and the 
LO results show progressively more strongly peaked
distributions.  A comparison with data should fairly easily test our
claim that the subtraction method is more correct.  
We also see that the 
number of 1-jet events and the number of 2-jets events predicted by our
subtraction method are of the same order, while the $k_t$-cut method and 
LO predict more dominance of 1-jet events\footnote{The number of 
events is roughly proportional to the area under the curve 
divided by the number of jets in an event.}. 
Also note that, in Fig.\ \ref{fig:smallxeta}, the peaks of the 
curves are not at the PM position $\hat{\eta} = 0 $. Even though the
peak corresponds to a smeared out $\delta(\hat\eta)$, the peak has shifted 
to $\hat\eta$ around $1/2$.  
The reason is that showering and hadronization of the jet changes a
massless on-shell quark into a massive quark, and the kinematics of
2-body collisions shows that this moves the rapidity of the quark
towards the proton's rapidity.  Since
the proton is chosen to move in the $+z$ direction, this means that
the rapidity of the jet is biased to a positive value relative to the
simple parton model value.

%==========================================================
\subsection{Cross Section at large $\xbj$}

\FIGURE{
\centering
\includegraphics*[1in, 3.2in][7in,7.6in]{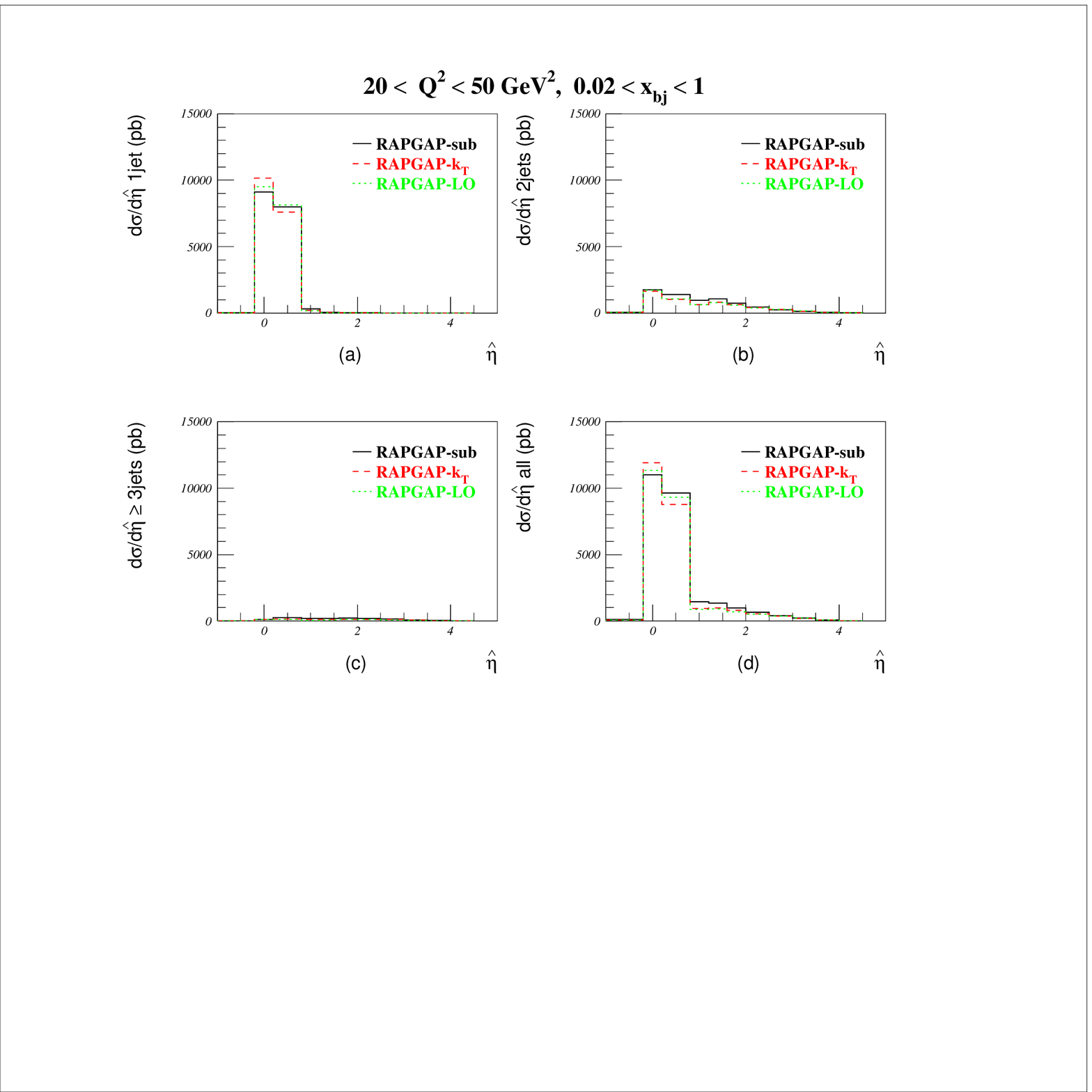}
\caption{Rapidity distribution of 1-jet inclusive cross section for 
$ 0.02 < \xbj < 1 $. The coding of the curves is the same as in Fig.\
  \ref{fig:smallxeta}.}
\label{fig:largexeta}
}

In Figs.\ \ref{fig:largexeta} and \ref{fig:largex} we show the results 
for large $\xbj$.  
The parameters are the same as before, except that we choose to
analyze events with $0.02 < \xbj < 1$. 
In contrast to the small $\xbj$ results, the three different methods give 
quite similar results, all peaked near $\hat\eta=0$ and $x_t=1$.  
In addition, most of the events are 1-jet events.
This is because gluon distribution is now much smaller, and the LO
process dominates.

\FIGURE{
\centering
\includegraphics*[1in, 3.2in][7in,7.6in]{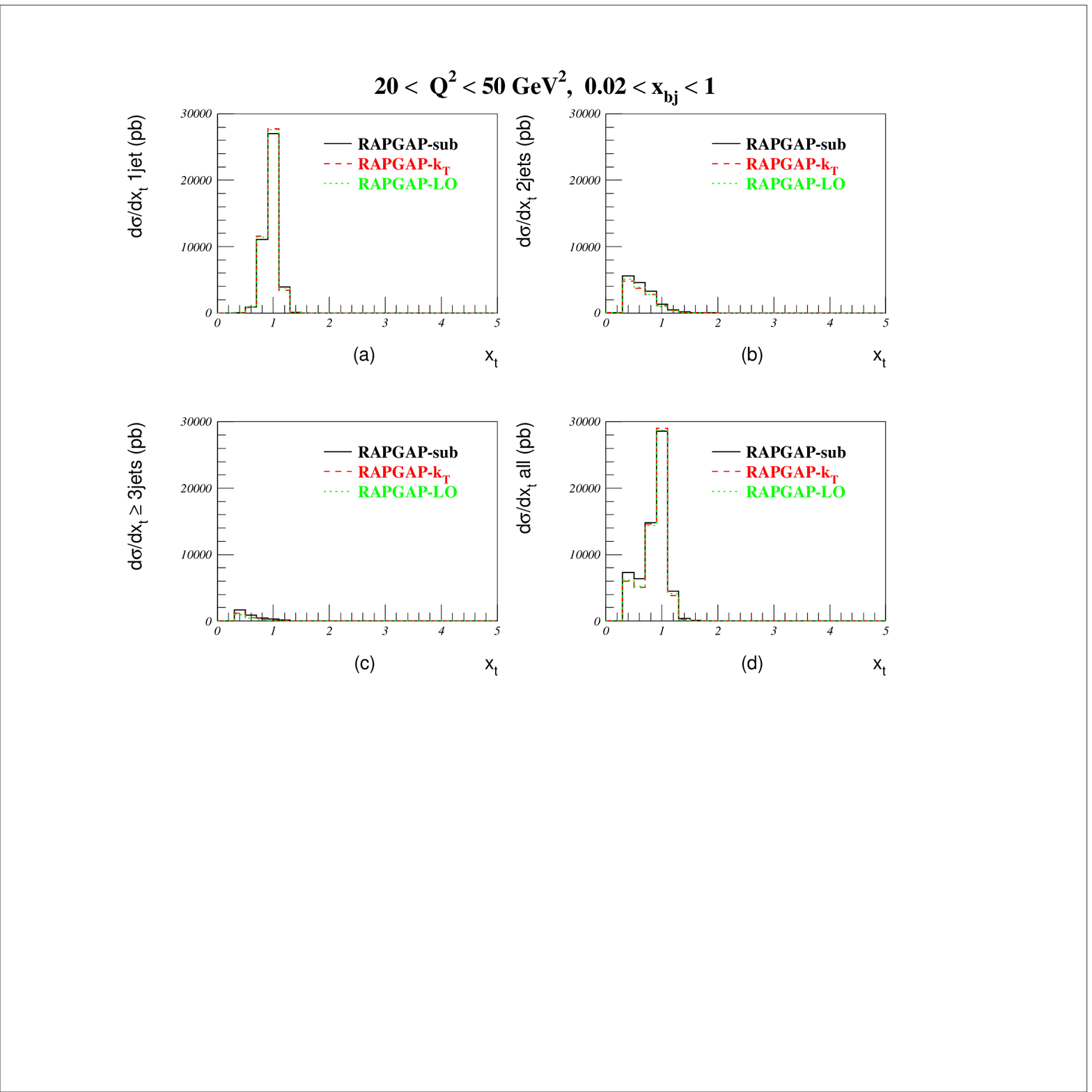}
\caption{Transverse momentum distribution of 1-jet inclusive cross
  section for $ 0.02 < \xbj < 1 $. 
  The coding of the curves is the same as in Fig.\ \ref{fig:smallxeta}.}
\label{fig:largex}
}
 
%----------------------------------------------------------
\section{Applications and Future Work}
\label{sec:concl}

We have implemented our subtraction method \cite{JC} in RAPGAP 
for DIS processes. The subtraction method is designed to give 
perturbative QCD predictions for the hard scattering with a correct 
treatment of parton kinematics.  We chose to calculate an infra-red
sensitive observable $d^2\sigma/dx_t d\hat{\eta}$ using three 
different methods, and to perform the calculation in a region where
the NLO correction is large because of a large gluon density.  We see 
large differences between the subtraction method and 
the $k_t$-cut method, thereby demonstrating the importance of the
treatment of parton kinematics.

There is no experimental analysis on $d^2\sigma/dx_td\hat{\eta}$ available 
yet, current analyses \cite{h1,zeus} being in the Breit frame. Since
our subtraction method, combined with the parton shower, is intended
to give
more accurate perturbative QCD predictions for infrared sensitive
cross sections,  
it would be interesting to do the experimental analysis of the jet cross
section in the laboratory frame relative to the position of
parton-model jet, and to
compare it with the MC calculations using our subtraction method. 

In the future we would like to work on two directions. One is to implement the 
subtraction method for diffractive DIS where there is also a large
gluon distribution.  Another is to work on the quark-induced NLO
subprocess in order 
to get a complete NLO MC event generator.

%==========================================================
\acknowledgments 
We would like to thank Professor Hannes Jung for his generous help,
various discussion and helpful suggestions. This work was supported in
part by the U.S.\ Department of Energy under grant number
DE-FG02-90ER-40577.

%==========================================================

\end{document}